\documentclass[12pt]{article}

\newcommand{\beq}{\begin{equation}}
\newcommand{\beqa}{\begin{eqnarray}}
\newcommand{\bega}{\begin{array}}
\newcommand{\ea}{\end{array}}

\newcommand{\eeq}{\end{equation}}
\newcommand{\eeqa}{\end{eqnarray}}
\newcommand{\p}{\partial}
\newcommand{\N}{{\hat N}}

\newcommand{\bone}{\mbox{\bf 1}}
\newcommand{\tr}{\mbox{tr}}

\newcommand{\half}{\mbox{$1\over2$}}

\newcommand{\OO}{\mbox{$\cal O$}}

\newcommand{\s}{\sigma}

\def \four{
{1\ov 4}
}

\def \ci {\cite}

\def \foot {\footnote}
\def \bi{\bibitem}
\def \tr {{\rm tr}}
\def \ha {{1 \over 2}}

\def\be{\begin{equation}}
\def\ee{\end{equation}}
\def\ba{\begin{eqnarray}}
\def\ea{\end{eqnarray}}
\newcommand{\bra}[1]{\mbox{$\langle #1 |$}}
\newcommand{\ket}[1]{\mbox{$| #1 \rangle$}}

\def\p{\partial}

\def\N{{\bf N}}

\def \d {\partial}

\def \ci{\cite}
\def \N {{\cal N}}

\def \adss{$AdS_5 \times S^5$\ }

\def \ov {\over}
\def \s{\sigma}

\def \ha {{1 \over 2}}

\def \la{\label}

\def\foot{\footnote}

\def\p{\partial}
\def\del{\partial}

\def\N{{\bf N}}

\def\tr{{\rm  tr}}

\def \lan  {\langle}
\def \ran  {\rangle}

\def \vn {\vec n} 
\def \cL{{\cal L}}
\def \rO {{\rm O}}
\def \roO {O}

\newcommand{\rf}[1]{(\ref{#1})}

\renewcommand{\thefootnote}{\fnsymbol{footnote}}
\def\appendix#1{
  \addtocounter{section}{1}
  \setcounter{equation}{0}
  \renewcommand{\thesection}{\Alph{section}}
  \section*{Appendix \thesection\protect\indent \parbox[t]{11.15cm}
  {#1} }
  \addcontentsline{toc}{section}{Appendix \thesection\ \ \ #1}
  }

\begin{document}
\def \inti { \int^{2\pi}_0 {d\s \ov 2\pi} }

\vskip-1pt
\hfill {\tt hep-th/0404215}
\vskip-1pt
\hfill BRX TH-540
\vskip0.2truecm
\begin{center}
\vskip 0.2truecm {\Large\bf
Towards the exact dilatation operator
\\ \vskip 0.2truecm
of ${\cal N}=4$ super Yang-Mills theory
}
\vskip 1.2truecm
{\bf
A.V. Ryzhov$^{1,}$\footnote{E-mail: ryzhovav@brandeis.edu}
and A.A. Tseytlin$^{2,}$\footnote{Also at Imperial College London
and  Lebedev  Institute, Moscow}\\
\vskip 0.4truecm
$^{1}$
{\it  Department of Physics,  Brandeis University\\
Waltham, MA 02454, USA}\\
\vskip .2truecm
$^{2}$ {\it Department of Physics,
The Ohio State University\\
Columbus, OH 43210-1106, USA}\\
 }
\end{center}
\vskip 0.5truecm
\vskip 0.2truecm \noindent\centerline{\bf Abstract}
\vskip .2truecm

We investigate  the structure of
the  dilatation operator $D$ of planar ${\cal N}=4$ SYM 
in the sector of single trace operators built out
of two chiral combinations of the 6 scalars.
Previous results at low orders in `t Hooft coupling $\lambda$ suggest that
$D$ has a form of an $SU(2)$ spin chain Hamiltonian with
long range  multiple spin  interactions.
Instead of the usual perturbative  expansion in powers of $\lambda$,
we split  $D$ into  parts $D^{(n)}$ according to the number $n$
of independent pairwise interactions between
spins at different sites. We determine the coefficients of spin-spin
interaction terms in  $D^{(1)}$ by  imposing the condition of  regularity
of the  BMN-type scaling limit. 
For long spin chains, 
these coefficients turn out to be expressible in terms of
hypergeometric functions of $\lambda$, which have  regular expansions  at
both small and large values of $\lambda$. This suggest that anomalous
dimensions of  ``long''  operators  in the two-scalar sector 
should generically scale  as $\sqrt{ \lambda}$ at large  $\lambda$,
i.e. in the same way as energies
of semiclassical states in dual $AdS_5 \times S^5$ string theory.

\def \aa {{\rm a}}
\def \tl {{\tilde \l}}
\def \N {{\cal N}}
\def \tr {{\rm tr \ }}
\newpage

\renewcommand{\thefootnote}{\arabic{footnote}}
\setcounter{footnote}{0}

\vskip 0.5truecm
\vskip 0.5truecm
\vskip 0.5truecm

\def \l {\lambda}
\def \D{{\cal D}}
\def \inn {Inozemtsev\ }
\def \G {\Gamma}

\setcounter{equation}{0}
\setcounter{footnote}{0}
\setcounter{section}{0}

\section{Introduction}
\label{intro}

The $\N=4$ supersymmetric  $SU(N)$ Yang-Mills theory
\ci{gko}  is
the  basic example of a  4-d conformal theory  \ci{fin}.
It  is actually a family of CFT's  parametrized  by
$N$ and the `t Hooft coupling $\l= g^2_{\rm YM} N$.
To solve a CFT  amounts,
at least, to being able to compute  dimensions of
local gauge invariant conformal
operators  as functions of the parameters.
This  problem should simplify in the planar limit of
$N \to \infty$, $\l$ fixed.
In this limit the AdS/CFT duality conjecture  \ci{mal}
suggests that  conformal dimensions should
be smooth functions of $\l$,
and have regular expansions
at both large and small $\l$.

Important progress towards this nontrivial goal of 
understanding how anomalous dimensions depend on $\l$
was recently made by concentrating on states with large quantum numbers 
(see, in particular,
\ci{bmn,gkp,gro,sz,ft1,rus,mina,mz1,ft2,bmsz,ft4,bfst,as,emz,kru,serb,kmmz,krt}).
For some non-BPS states  there are new expansion limits
(like large $J$, fixed $\tl \equiv  {\l \over J^2}$
for $S^5$-rotating pointlike  \ci{bmn}  and
multispin \ci{ft2} string states) 
where one can directly compare
perturbative SYM anomalous dimensions 
to semiclassical string results.

One would obviously like to go
beyond the restriction to long and/or scalar-only
operators and compute,
e.g., the exact dimension of the Konishi scalar
$\tr (\Phi^*_i \Phi_i)$, or the coefficient $f(\l)$ of the
$\ln S$ term in the  anomalous dimension of the minimal
twist operators 
such as $\tr (\Phi^*_i  D^S \Phi_i)$
\ci{gkp,ft1,kot}.
The main obstacle is our
lack of tools for obtaining exact all-order results 
on either gauge theory ($\sum c_n \l^n$) or
string theory ($\sum  {b_n\ov {(\sqrt \l)}^n}$) side.
One potentially fruitful idea of how to go beyond the first few orders
in SYM perturbation theory 
is to try
to determine 
the exact structure of the 
dilatation operator $D$ by imposing additional
conditions (like superconformal symmetry,
BMN limit, integrability, etc.,  as in
\ci{bks,beif,bei})
 implied by the expected
 correspondence with \adss string theory.
Having found the resulting
anomalous dimensions as functions of $\l$, 
one may then be able to 
see if
they 
admit a regular expansion
not only at small  but also at large $\l$.

This is the approach we would like to explore below
using as an input the condition
of  regularity of  BMN-type scaling limit in the form
suggested in  \ci{kru} and  further clarified  in \ci{krt}.
We shall concentrate on the planar  $SU(2)$ sector of  single trace
SYM operators
built out of chiral
combinations $X$ and $Z$ of two the 6 SYM adjoint scalars, i.e.
$\tr ( X...X Z...Z X...)$  with canonical dimension $L$.
This sector is closed under renormalization \ci{bks}.
The eigen-operators of $D$ with 
$J_1$ $Z$'s and $J_2$ $X$'s
(so $L=J_1+J_2$)  should be dual to string states
with two components of the $SO(6)$ spin \ci{bmn,ft2}.
On general grounds, the SYM dilatation operator computed in the planar
limit should be a series in $\l$
\be \la{dod}
 D=\sum_{r=0}^\infty { \l^r \ov (4 \pi)^{2r}}   D_{2r}  \ . \ee
 Let us review what is known already  about the structure
 of $D_{2r}$.

Restricting $D$ to planar graphs   suggests  that
$D_{2r}$ should be given by  local sums over  sites
$a=1,...,L$ with $Z$ and $X$ interpreted as a spin ``up''
and spin ``down'' state of a periodic ($a+L\equiv a$)
spin chain \ci{mz1,bks}
for which $D$ is the Hamiltonian, 
\be\la{loc}
 D_{2r} =   \sum^{L}_{a=1} \D_{2r}(a) \ , \ \ \ \ \ \
 \ \   \D_0 = \bone \ .
\ee
The one-loop term $\D_2$ turns out to be equivalent to the
Hamiltonian
of the ferromagnetic  XXX$_{\ha }$  Heisenberg spin chain \ci{mz1},
\be\la{chg}
\D_2= 2Q_{a,a+1} \ , \ \ \ee
\be \la{qq}
 Q_{a,b}\equiv \bone  - P_{a,b}  \ , \ \ \ \ \ {\rm i.e.} \ \ \ \ \ \ \
  Q_{a,b}=
  \ha (\bone - \vec \s_a \cdot \vec \s_b)
\ , \ee
where $P_{a,b}$ is the   permutation operator
and $\vec \s_a$ are the Pauli matrices acting on the spin
state at site $a$.\foot{For $a=b$ one should set $Q_{a,a}=0$; note that
$P^2=\bone$
and $\ha Q$ is a projector.}
The two-loop  term $\D_4$ was found to be  \ci{bks}
\be \la{iy}
\D_4=  2(- 4 Q_{a,a+1}  + Q_{a,a+2})   \ , \ee
while the expression for the 3-loop term $\D_6$
conjectured in \ci{bks} on the basis of integrability
\beqa
\D_6 &=& 4( 15Q_{a,a+1} - 6  Q_{a,a+2}  +  Q_{a,a+3 }) 
\nonumber\\
\la{ity}
&& +  4 (Q_{a,a+2} Q_{a+1,a+3}   -  Q_{a,a+3} Q_{a+1,a+2})
   \ , 
\eeqa
was shown in \ci{bei} to be  uniquely fixed
by
the superconformal symmetry algebra, constraints coming from the
structure of Feynman graphs and the correct BMN limit.%
\foot{The same expression was found in a closely related
context of SYM matrix model \ci{plefka}.
Also, the 3-loop anomalous dimension of (a descendant) of the Konishi
operator found in \ci{bks} from the above form of  $D_6$
received a remarkable indirect confirmation
in a recent computation of anomalous dimension
in twist 2 sector \ci{kot}
which also contains a  descendant of  the Konishi operator 
(N. Beisert and M. Staudacher, private communication).}
Finally, there is also a  proposal \ci{beif}
for the 4-loop term   $\D_8$ based on assuming
integrability \ci{bks} and the BMN scaling.
Written in terms of factorized permutations as in \ci{serb}
it reads
\beqa
\D_8 &=&
 10(- 56 Q_{a, a+1} + 28 Q_{a, a+2} - 8 Q_{a, a+3} +  Q_{a, a+4})
\nonumber\\&&
\la{nex}
 +  \frac{2}{3}\bigg[
 ( 421 Q_{a,a+1}Q_{a+2,a+3}+ 986 Q_{a,a+3} Q_{a+1,a+2}- 183
Q_{a,a+2}Q_{a+1,a+3})
\nonumber\\&&
+  8( Q_{a,a+3}Q_{a+2,a+4}+Q_{a,a+2}Q_{a+1,a+4}-
Q_{a,a+4}Q_{a+2,a+3}-Q_{a,a+4}Q_{a+1,a+2} 
\nonumber\\&&
-\ Q_{a,a+4}Q_{a+1,a+3}+Q_{a,a+3}Q_{a+1,a+4})\bigg]\ . 
\eeqa
Generalizing  the above expressions for
$r=2,3,4$ it is then natural to expect that  generic $r$-loop
term in  \rf{loc} will  contain a term  linear in $Q_{a,b}$,
a term quadratic in $Q_{a,b}$, and so on:
\beqa \la{stru}
\D_{2r} &=& \D^{(1)}_{2r} + \D^{(2)}_{2r} + ...
\ , \hspace{2em} \ \ \ \ \ \ \
\D^{(n)}_{2r} \sim  \sum Q^n  \ ,
\\
\la{ru}
&&
\D^{(1)}_{2r} = 2\sum_{c=1}^r \aa_{r,c} Q_{a,a+c}  \ .
\eeqa
At order $r$  there can be
at most $r$ spin-spin interactions in $\D^{(1)}_{2r}$ \ci{gro,bks}.
$Q^n$ in $\D^{(n)}_{2r}$ stands for products of independent projectors,
i.e.  with all indices corresponding to different sites as in
\rf{ity} and \rf{nex}.
The above explicit  expressions \rf{chg}, \rf{iy}, \rf{ity}, \rf{nex}
for $D_2,...,D_8$  imply that
for $r\leq 4$ the coefficients
$\aa_{r,c}$  are ($c=1,2,...,r$)
\be\la{ah}
\aa_{1,1} = 1\ ; \ \ \  \aa_{2,c} = (-4,1)\ ; \ \ \
\aa_{3,c} = (30,-12,2) \ ; \ \ \
\aa_{4,c}= (-280,140,-40,5) \ . 
\ee
Then
\be \la{dee}
D=D_0 +  D^{(1)} + D^{(2)} + ... \ , \ \ \ \ \ \ \ \ \
D^{(1)}=  2\sum_{r=1}^\infty { \l^r \ov (4 \pi)^{2r}}
 \sum_{a=1}^L
 \sum_{c=1}^r \aa_{r,c} Q_{a,a+c} \ . \ee
Using the  periodicity of the chain ($Q_{a,b+L}=Q_{a,b}$, etc.)
 $D^{(1)}$ can be rewritten as
 \be \la{hih}
 D^{(1)}=   \sum_{a=1}^L
 \sum_{c=1}^{L-1} h_c(L,\l)\  Q_{a,a+c}
 \ . \ee
Our aim below will be to determine the general expression for the
coefficients $\aa_{r,c}$ and thus the functions $h_c(L,\l)$,
i.e. to find
the spin-spin (linear in $Q$)  part
of the exact dilatation operator $D$.


To find the coefficients $\aa_{r,c}$ in \rf{ru},
we will demand that the BMN-type scaling limit
\be \la{sca}
L\to \infty \ , \ \ \ \ \ \ \ \
\tl \equiv  {\l\ov L^2} ={\rm fixed}
\ee
of the  coherent-state
expectation value
of $\D^{(1)}$ \ci{kru,krt}
is well defined.
This  turns out to be
(nearly)  equivalent to the consistency  with  the   BMN
expression \ci{bmn,gro,sz}  for the anomalous dimensions
of the 2-impurity operators.
Imposing the  condition of agreement with the BMN square root formula
fixes one remaining free coefficient at each order in $r$.

Our approach is thus  
similar  to  the
 previous important investigations
of the constraints on the dilatation operator imposed by the BMN
limit \ci{gro,bks,beif,bei}.
The  new elements of the  present
discussion are  that
(i)  we  follow \ci{serb,krt} and
classify the structures  in $D$
in terms of independent interactions between sites as
in \rf{stru},\rf{dee}, i.e.
$D=D_0 +  \sum Q  + \sum QQ + ...$, and
(ii) we  resum the loop  expansion  to find
the coefficients in $D^{(1)} =  \sum Q  $ as
 explicit functions of $\l$ and study their strong-coupling limit.


The resulting $D^{(1)}$ \rf{hih}  
may be interpreted as a 
Hamiltonian
of a periodic spin chain with  long-range interactions.
One could  conjecture
that,  like in  some  known examples  \ci{sha,ino},
this spin chain  may be integrable;
and, furthermore,
the higher-order  terms $D^{(2)}, D^{(3)}, ...$
in  \rf{dee}
may be
effectively determined by $D^{(1)}$, e.g., expressed in
terms of higher conserved charges of the  chain.
This would then
determine the full 
$D$.
Remarkably, 
this is indeed true up
to  order
$\l^3$
\ci{serb}: 
the sum of one, two, and three-loop dilatation operators
can be viewed as a
part of  the Inozemtsev integrable spin chain \ci{ino},
with the $QQ$-terms in
$D_6$ in  \rf{ity} being proportional to a leading  term in the
$\l$-expansion  of a higher
conserved charge of the Inozemtsev chain.
At order $\l^4$ the Hamiltonian of the Inozemtsev chain does not,
however, agree with the BMN perturbative scaling \ci{serb}.
Here, instead of starting with the Inozemtsev chain,
we reverse the logic and determine which spin chain Hamiltonian
is actually consistent with  the BMN  limit,
leaving the issue of integrability open.\foot{Footnote 5 in \ci{serb}
points out, following \ci{ino}, 
 that the only integrable  spin chain with  spin-spin
interactions is the \inn   chain.  This implies
that higher-order terms $\D^{(n)}$
 may not be directly determined by $\D^{(1)}$, and only
 the whole $D$ may represent a Hamiltonian of an integrable
  spin chain.}

  For comparison, let us recall that the \inn Hamiltonian that
  interpolates between the Heisenberg and Haldane-Shastry
  spin chain Hamiltonians is given by \ci{ino}
  \be \la{ink}
  H= \sum_{a=1}^L \sum_{c=1}^{L-1} p_{c}(L,q)\  Q_{a,a+c}
 \ , \ee
 where
 $ p_{c}(L,q)$ is a double-periodic  Weierstrass function
 with periods $L$ and $q$
 \be p_{c}=
 {1\ov c^2} + \sum_{(m,n)\not=(0,0)}
\left[
{ 1 \ov ( c - m L - i n q)^2 }
 - { 1 \ov (  m L +  i n q)^2 }
 \right] \ . \ee
Note that $p_c= p_{L-c}$, so the sum in \rf{ink} may be restricted to 
$c \leq [L/2]$.
The limiting cases are   $\lim_{q\to \infty} p_c  = ({\pi \ov L})^2
  ( { 1 \ov \sin^2 { \pi c \ov L}} - { 1 \ov 3}) $
  (the Haldane-Shastry chain limit), and
  $\lim_{L\to \infty} p_c
  = ({\pi \ov q})^2
  ( { 1 \ov \sinh^2 { \pi c \ov q}} + { 1 \ov 3}) $
  ($q\to 0$ corresponds
   to  the infinite Heisenberg chain).
As was suggested in \ci{serb}, to relate $H$ in \rf{ink}
 to  the 3-loop  SYM dilatation operator
  one is to relate $q$ to $\l$ by
 $
 {\l \ov (4 \pi)^2} = \sum_{n =1}^\infty
  {  (4 \sinh^2 { \pi n \ov q})^{-1} }.
 $


The rest of this paper is organized as follows.
In section 2 we 
determine the coefficients in \rf{ru}
by first imposing the regularity of the scaling limit \rf{sca}
of the coherent-state expectation value of $D^{(1)}$
following \ci{kru,krt} (section 2.1) and then
checking consistency with the 2-impurity BMN spectrum
(section 2.2). In principle, we  could  fix all the coefficients
just by  imposing  the second (BMN) condition
but we believe the approach
of section 2.1 is more straightforward and
has its own conceptual merit, having  close connection
to string theory \ci{kru,krt}.

In section 3 we 
discuss how to sum up
the 't Hooft coupling expansion
of the  coefficients in $D^{(1)}$.
We 
first show that in the large $L$ limit
the coefficients of $Q_{a,a+c}$ can be expressed
in terms of 
hypergeometric functions
which smoothly interpolate  between  perturbative power series
at small $\l$  and $\sqrt \l$ growth at large $\l$.
We  
then comment on the  finite $L$ case,
in particular on the resulting contribution of $D^{(1)}$
to the exact anomalous dimension of the Konishi operator.

Section 4 
contains
some conclusions and 
a discussion of
open problems.
Some useful  relations  and definitions 
are
summarized in Appendix.

\setcounter{equation}{0}
\section{Determining the coefficients in  $D^{(1)}$}
\label{fixing D1}

Our first task will be to obtain the general expression 
for the coefficients $\aa_{r,c}$  in the linear in $Q_{a,a+c}$ part 
of the dilatation operator in \rf{loc},\rf{ru}. 
To do this we shall follow   \ci{kru,krt}
and consider the spin coherent state path integral
representation for  the  quantum mechanics 
of $D$ \rf{dod} as a generalized spin chain Hamiltonian.  
The corresponding action for a collection of unit 3-vectors 
$\vec n_a(t)$ at each of $L$ sites of the chain 
($\bra{n} \s^i_a\ket{n} = n^i_a$, \  $(n^i_a)^2=1, $\  $i=1,2,3$)
 will then be given by 
 \be \la{kh}
 S= \int dt\  \bigg[
   \sum^L_{a=1} \cL_{\rm WZ} (n_a) - \lan  D \ran   \bigg] \ ,
 \ \ \ \ \ \ \ \     \lan D \ran  = \bra{n} D \ket{n}\ , \ee
 where $\cL_{\rm WZ} (\vec n_a) = C_i (\vec n_a) \dot  n_a^i  $ 
 ensures the proper $SU(2)$  commutation relations if 
  one considers  $\vec n_a$ back as
spin   operators (see, e.g., \ci{fra}). 
Motivated by the explicit  results \rf{chg}--\rf{nex} where 
the leading (at small $\l$) coefficient of $Q_{a,a+c}$ is always positive,  
it is natural to assume that the spin chain in question is ferromagnetic.
\footnote{Then  the state with all spins ``up'' (represented by the operator $\tr Z^L$) 
is a true vacuum.
}
Then in the long spin chain limit $L \to \infty$ one 
may expect \ci{kru,krt} that the low energy excitations 
of the spin chain will be captured by the semiclassical dynamics 
of
\rf{kh}.  
The correspondence with string theory 
then suggests \ci{kru,krt} that $S$ should have a regular 
scaling limit \rf{sca}, or, more explicitly, 
that the low energy effective action for the system governed by (\ref{kh})
should have a well defined
continuum limit. 
To take the continuum limit one  may 
introduce a field  $\vec n(\s,t)$,
 $ 0 < \s \leq  2\pi, $  with $ \vec n_a(t)= \vec n({2\pi a \ov L},t )$, 
so that \rf{kh}  becomes 
\be \la{gh}
S\to   L  \int dt \inti \  \bigg[ C_i (\vec n) \dot  n_a^i - 
{\cal H} ( \del \vn,\del^2 \vn ,...;  \tl)  \big] \ .  
\ee
$\cal H $ which originated from $\langle D \rangle$ 
should be a regular function of 
the effective coupling $\tl$
and 
$\s$-derivatives of $\vec n(t, \s)$ 
in the limit $L \to \infty$, $\tl$ fixed
(with subleading $1\over L^n$ terms omitted). 
Quantum corrections are then suppressed because of the 
large prefactor $L$ in 
front of the action. 

Writing $D$ in terms of  factorized permutation operators  as in 
\rf{iy}--\rf{nex} one observes that since the $Q_{a,b}$ in \rf{qq} 
satisfy 
\be 
\la{rty}
\bra{n}Q_{a,b} \ket{n} = \ha ( 1 - \vn_a \cdot \vn_b) 
= \four  ( \vec n_a - \vec n_b)^2 \ , 
\ee
$\bra{n}\D^{(1)}\ket{n}$ in \rf{stru} 
contains terms 
{\it quadratic} in $\vn$ 
(but all orders in  derivatives);  
$\bra{n}\D^{(2)}\ket{n}$, terms {\it quartic} in $n$, etc. 
The approximation  that distinguishes 
$ \D^{(1)}$ from all higher $ \D^{(k)}$ in \rf{stru} 
is the one in which one keeps only small 
fluctuations of $\vec n(t,\s)$ near its (``all-spins-up'') ground-state  
value $\vec n_0 = (0,0,1)$. Then 
$\vec n = \vec n_0 + \vec {\delta n}$, where $|\vec {\delta n}| \ll 1$, 
so that higher powers of the fluctuating field $\vec {\delta n}$
are suppressed, regardless the number of spatial derivatives
acting on them. 
Such configurations 
correspond to semiclassical spinning string  states 
with  $J_1 \gg J_2$, and are close to a  single-spin BPS state.
They should indeed represent semiclassical or 
coherent-state analogs of few-impurity BMN states, 
having  the same  BMN energy-spin relation which is  
indeed reproduced in the limit  $J_1 \gg J_2$  \ci{ft1,ft2,ft4,bfst}
by the classical two-spin string solutions.

Let us therefore consider this BMN-type approximation,   
concentrating on the part of $\bra{n} D \ket{n}$ which is quadratic in $n$, 
i.e. on  $\bra{n} D^{(1)}  \ket{n}$, 
and demand 
that the continuum 
version of $\bra{n} D^{(1)} \ket{n}$ 
have a regular
scaling limit \rf{sca}.
As was  shown  in \ci{kru,krt}, this condition is
indeed satisfied at $r=1,2,3$ loop orders, i.e. for \rf{chg},\rf{iy}
and \rf{ity} (in   
general, this  should  be a consequence of the 
supersymmetry of the underlying  SYM theory which
restricts the structure of $D$).
One finds that 
the coefficients 
in the order $Q$-terms in \rf{iy}--\rf{nex} 
are such that all lower than $r$-derivative terms in the continuum limit of 
$\bra{n} \D^{(1)}_r  \ket{n}$ cancel out. If they did not, 
the limit $L \to \infty$ would be singular, as there would 
be a disbalance between the powers of $\l$ and the powers of $L$.
Explicitly, one gets  for $r=1,2,3,4$ using \rf{ah}
(after integrating by parts) 
\beqa \la{lety}
{ \l^r \ov (4 \pi)^{2r}} \lan\D_{2r}^{(1)}\ran 
= { \l^r \ov 2 (4 \pi)^{2r}} 
\sum_{c=1}^r \aa_{r,c} (\vn_a - \vn_{a+c})^2 
&\to&  
 d_r \tl^r
\left[ (\del^r \vn)^2 + \OO \left({\d^{2r+2}  \ov L^2} \right)
\right],  
\eeqa
where
\be d_1 = { 1 \ov 8}\ ,  \ \ \ \ \ 
d_2 = - { 1 \ov 32}\ , \ \ \ \ \ 
d_3 ={ 1 \ov 64} \ , \ \ \ \ \ 
d_4 = -{ 5 \ov  512} \ . \ee 
The sum  over  $a$ in \rf{loc} becomes an  integral over $\s$
as in \rf{gh}, 
and we find
$\sum_{a=1}^L \D_{2r} (\s_a) \to L \left( \int_0^{2 \pi} {d\s \over 2 \pi}
\D_{2r} (\s) + \OO({1\ov L}) \right)$.

\subsection{Regularity of the continuum limit}
\label{good lim}

Let us now demand that the same pattern of cancellations 
\rf{lety} 
should persist to all orders in $\l$-expansion.   

Taking the continuum limit in \rf{lety} one may use the Taylor expansion 
to show that, up to a total derivative, 
\beqa
\label{eq: n n}
(\vn_a - \vn_{a+c})^2 
&=&
2  \sum_{m=1}^\infty
{(-1)^{m-1} (2 \pi c)^{2 m} \over (2 m)! L^{2 m}} (\p^m \vn)^2
+ \p (...)\ . 
\eeqa
Then 
\beqa
\label{cont}
{ \l^r \ov (4 \pi)^{2r}} \langle \D_{2r}^{(1)} \rangle &\to&
 \l^r \sum_{m=1}^\infty { d_{rm} \ov L^{2m} } (\p^m \vn)^2 \ , 
\eeqa
\be \la{vee}
d_{rm} =  {(-1)^{m-1} (2 \pi)^{2 m}\over (4 \pi)^{2r} (2 m)! } 
\sum_{c=1}^r c^{2 m} \aa_{r,c} \ . 
\ee 
To make sure that the limit \rf{sca} of $D^{(1)}$ is well defined, 
$\langle \D_{2r}^{(1)} \rangle$ should scale as $1\ov L^{2r}$ so that 
$\l^r \langle \D_{2r}^{(1)} \rangle \sim \tl^r= {\l^r\ov L^{2r}}$, 
up to subleading $ \OO({1\ov L}) $ terms. 
This implies that $d_{r1},...,d_{r,r-1}$ must vanish, 
i.e.  that the coefficients $\aa_{r,c}$ must satisfy
\beqa
\label{BMN 0}
\sum_{c=1}^r c^{2 m} \aa_{r,c} = 0\ , 
\quad\quad\ \ \
\mbox{ for }\ \  0 < m < r \ .  
\eeqa
This gives $(r-1)$ equations for $r$ unknowns.
The  coefficients $\aa_{r,c}$ with  $c = 2, ... , r-1$ 
are then uniquely determined in terms of  
 a single multiplicative constant $\aa_{r,r}$  from
\beqa
\label{BMNeq}
\left(\matrix{ 
1 & 2^{2} & \cdots & (r-1)^{2} \cr 
1 & 2^{4} & \cdots & (r-1)^{4} \cr 
& \vdots & \vdots & \vdots \cr
1 & 2^{2(r-1)} & \cdots & (r-1)^{2(r-1)} 
} \right)
\left(\matrix{ \aa_{r,1}  \cr \aa_{r,2} \cr \vdots \cr \aa_{r,r-1} \cr } \right)
= - \aa_{r,r}
\left(\matrix{ r^2 \cr r^{4} \cr \vdots \cr r^{2(r-1)} \cr } \right)\ . 
\eeqa
The matrix in (\ref{BMNeq})
is  invertible for any $r$,
and one finds\foot{The same relation appeared in related context of
BMN limit  in the first reference in  \ci{gro}.
There the  authors computed    contributions to the scaling dimensions
of the BMN operators by  analyzing  the relevant  Feynman diagrams.
Here instead we  determine  the part of the full dilatation operator
which, in particular,  computes  the dimensions of BMN operators,
 but which can also be
applied systematically to 
 general operators made out  of scalars $X$
and $Z$.} 
\beqa
\label{BMNsol}
{ \aa_{r,c}  } =  {  (-1)^{r-c}(2r)! \over (r-c)! (r+c)!}\  \aa_{r,r}\ \ , 
\quad\quad\qquad 
c = 1, ..., r-1\ .
\eeqa
This generalizes \rf{ah} to all values of $r$.
Given (\ref{BMNsol}), the first non-vanishing coefficient 
$v_r$ in \rf{vee} becomes 
\be \la{hhh}
d_{rr} \equiv d_r =    {(-1)^{r-1} \over 2^{2 r} (2 r)! } 
 \sum_{c=1}^r c^{2 r} \aa_{r,c} 
= {(-1)^{r-1} \ov  2^{2 r+1 } }   \aa_{r,r}   \ ,  \ee
and
the $r$-loop contribution to 
the  expectation value of 
$D^{(1)}$ takes the form (cf. \rf{lety}) 
\beqa
\label{kl}
{\lambda^r \over (4 \pi)^{2 r}}
\langle D_{2r}^{(1)} \rangle &\to&
L \left[  d_r \tl^r   \inti (\p^r \vn)^2 
 + \OO({1\ov L}) \right]
\eeqa
What remains is 
to find the 
values 
of $\aa_{r,r}$ generalizing 
\rf{ah}.
This can be done by 
analyzing the spectrum of BMN operators.

\subsection{Constraints from the BMN limit}
\label{BMN}

Let us  consider the two-impurity BMN operators of the form
\beqa
\label{BMN ops}
\rO_n^{\rm BMN}  &=& 
{1\over \sqrt{J+1}} \sum_{p=0}^J \cos \left[{\pi n (2 p + 1)\over J+1} \right] \; \tr (X Z^p X Z^{J-p}) \ . 
\eeqa
These operators are multiplicatively renormalized for any $J$, 
and so are eigenstates of $D$ 
(see \ci{bebmn} and refs. there).
Here the total number of fields, i.e. the length of the spin 
chain, is $L=J+2$, 
with  $J_1\equiv J$ 
fields  
$Z$ and $J_2=2$ 
fields  
$X$.
Anomalous dimensions of the BMN operators can 
be computed in both string theory and gauge theory \cite{bmn,sz}
in the large $J$, fixed ${ \l \ov J^2}$ limit, 
and one finds
\be
\label{BMN dims}
\Delta_{\rm BMN} = J + 2 \sqrt{1 + {\lambda\ov J^2}  n^2}
+  \OO({1\ov J})\ 
 .
\ee
Note that since $L=J+2$, this  BMN limit is essentially the same as 
the scaling limit \rf{sca} discussed above, with 
$\tl$ being the same, up to subleading $1\ov J$ terms, 
as  $ {\l \ov J^2}  $ (which is usually denoted as $\l'$).  
To reproduce \rf{BMN dims} (i.e. the coefficients in the 
expansion of the square root in powers of $\tl$) by acting with the 
dilatation operator (\ref{dod}) on $\rO^{\rm BMN}_n$, we 
should
require that 
\beqa
\label{D BMN}
D_{2r} \rO^{\rm BMN}_n = 
4 {(-1)^{r-1} \Gamma(2r-1) \over \Gamma(r) \Gamma(r+1)}
\left( {2 \pi n \over J} \right)^{2r} \rO^{\rm BMN}_n + 
\OO({1\ov J^{2r+1}})\ . 
\eeqa
To  evaluate $D_{2r} \rO^{\rm BMN}_n$ explicitly let us note  
since we are interested in the large $J$ limit,
we can ignore what happens near the ends of the spin chain. 
Then  for generic values of $p$ in the sum \rf{BMN ops} 
we find that the permutation operators $P_{a,b}= \bone - Q_{a,b}$ 
act as follows ($c$ can be positive or negative)
\beqa
\label{P on BMN}
P_{1,1 + c} \roO_p = P_{p+2,p+2-c}\roO_p = \roO_{p-c}\ , \ \ \ \ \ 
\ \  \roO_{p} \equiv \tr (X Z^p X Z^{J-p}) \ , 
\eeqa
while for all other labels $a,b$ we get 
 $P_{a,b} \roO_p  =\roO_p$.
Then \beqa
\label{Q on BMN}
\sum_{a=1}^{J+2} Q_{a,a+c}\ \roO_p = 
2 (\roO_p - \roO_{p-c}) + 2 (\roO_p - \roO_{p+c}) \ , 
\eeqa
where there are four nonzero combinations, 
from $Q_{1,1 \pm |c|}$ and  from $Q_{p+2,p+2 \pm |c|}$.
Other terms $ \D^{(n)}_{2r}$ in $D_{2r}$ \rf{stru} 
containing two and more products of $Q$'s 
(which appear starting with 3-loop term \rf{ity}), i.e. 
 $Q_{a,b}...Q_{c,d}$
with all labels  $a,b...c,d$ distinct,
annihilate $\roO_p$ unless we are near the ends
of the chain, 
i.e. they do not contribute to (\ref{D BMN})
in the large $J$ limit. 
Ignoring these higher order
terms is equivalent to the usual 
dilute gas approximation which applies 
when the number of impurities is small and the spin chain is very long.
Thus it is the $D^{(1)}$ part 
of the full dilatation operator \rf{dee} that is 
responsible for the anomalous dimensions of the operators
$\rO_n^{\rm BMN}$ in the BMN limit, 
\beqa
\label{Dfull BMN}
(D - D_0 ) \rO_n^{\rm BMN} = D^{(1)} \rO_n^{\rm BMN} 
+ ...\  , 
\eeqa
where dots stand for  subleading $1\ov J$ terms. 
Using \rf{Q on BMN} we find that to the leading order in $1\ov J$,
\beqa
\label{DonBMN}
D^{(1)}_{2r} \rO_n^{\rm BMN}
&=&
\sum_{c=1}^{r} \aa_{r,c} 
{4 \over \sqrt{J+1}} \sum_{p=0}^J \cos \left[{\pi n 
(2 p + 1)\over J+1} \right]\ 
(2 \roO_p - \roO_{p-c} - \roO_{p+c})
\nonumber\\
&=&
8 \sum_{m=1}^\infty
{ (-1)^{m-1} \over (2 m)! }
\left({2 \pi n \over J} \right)^{2m}
\sum_{c=1}^{r} c^{2m} \aa_{r,c} \ 
\rO_n^{\rm BMN} \ . 
\eeqa
To arrive at this expression we  expanded the cosines
at large $J$.  
The sum $ \sum_{c=1}^{r} c^{2m} \aa_{r,c}$  in (\ref{DonBMN}) 
is precisely  the same  as in (\ref{cont}),\rf{vee}.
Using the results (\ref{BMN 0}) and (\ref{hhh}) for $\aa_{r,c}$,
\beqa
\label{D2onBMN}
D_{2r} \rO_n^{\rm BMN}
&=&
4(-1)^{r-1} \aa_{r,r}  \left({2 \pi n \over J} \right)^{2r}
\ \rO_n^{\rm BMN}+ \OO({1\ov J^{2r+1}})
\nonumber\\&=&
(-1)^{r-1}
\aa_{r,r} 
\left( D_2 \right)^r\
\rO_n^{\rm BMN}
+ \OO({1\ov J^{2r+1}})
\ , \eeqa
so 
to match the BMN expression (\ref{D BMN}) we must have
\beqa
\label{aBMN}
\aa_{r,r} 
={\Gamma(2r-1) \over \Gamma(r)\ \Gamma(r+1)}
= {2^{2r-3} \G(r- \ha) \ov \sqrt \pi \ \G(r+1) } \ . 
\eeqa
For $r=1,2,3,4$ this gives  $\aa_{r,r}= 1, 1, 2, 5$
 as in \rf{ah} (note that the limit of $r\to 1$ 
of the the second expression in \rf{aBMN}
is well defined).
Combining \rf{BMNsol} with \rf{aBMN} we 
finally conclude that  
\be \la{fin}
{ \aa_{r,c}  } =  
{   (-1)^{r-c}\  \Gamma( 2r+1)\   \Gamma(2r-1) \over 
\Gamma( r-c+1) 
 \ \Gamma(r+c+1 )  \ \Gamma(r)\ \Gamma(r+1)   }  \ .
\ee
Note that, as required (cf. \rf{ru}), 
 $\aa_{r,c}=0$  for $ r < c$ (for integer $r$ and $c$).

\setcounter{equation}{0}
\section{Summing up: structure of $D^{(1)}$ to all orders in $\l$}
\label{sum D1}

Let us now try to draw some conclusions  about the exact 
structure of the ``spin-spin'' part of the dilatation 
operator $D^{(1)}$ in \rf{dee} to all orders in $\l$.
First, let us   observe 
 that the value of $\aa_{r,r}$ found in \rf{aBMN}
implies that summing \rf{kl} over $r$ with $d_r$ in  \rf{hhh}
gives a very simple formula for the quadratic in $\vn$ 
(``small fluctuation'' or ``BMN'') 
part of the  coherent state effective Hamiltonian  in \rf{gh}:
\beqa
\label{D KRT}
\langle D^{(1)} \rangle =
\sum_{r=1}^\infty 
{\lambda^r \over (4 \pi)^{2r} } \langle D_{2r}^{(1)} \rangle 
&\to &
L  \inti
\left[ \  
{1\over 4}\  \vn \left( \sqrt{1 -\ \tl\ \p^2} - 1 \right) \vn
+ \OO({1\ov L}) \right] \ . 
\eeqa
Remarkably, this is the same expression 
 that follows 
from the classical \adss string sigma model action 
expanded in the limit \rf{sca} (eq. (2.90) in \ci{krt}).
There is a closely related square root 
formula (see eq. (3.25) in \ci{krt}) that expresses  
 $D^{(1)}$ as a function of  $D_2$ in the dilute gas 
approximation:
\beqa
\label{D1}
D^{(1)} =
 \left( \sqrt{1 + 2{\lambda \over (4 \pi)^2} D_2} - 1 
\right)^{(1)}\ .
\eeqa
The superscript ``${(1)}$" 
in the right hand side means that one  should drop all terms
 with higher than first power of independent 
$Q_{a,b}$'s (written in the  factorized form) 
in the products  of $D_2$.
This  relation should be understood  in the sense 
of equality of the $\vn^2$ terms in the  coherent-state expectation values 
of the two sides.\foot{
If we wanted to apply $D^{(1)}$ to a single trace operator, 
the compact expression in the right hand side of (\ref{D1}) 
would be of limited advantage. 
Multiplying $D_2$'s 
and taking the single $Q_{a,b}$ part do not commute,
and we would have to explicitly compute all powers of $D_2$ and then
apply the ``${(1)}$-operation" to them
(like it  was done in  Appendix C  of \cite{krt} 
where connected expectation values of products 
of some operators  were computed).}

Next,  let us  substitute the values 
\rf{fin} for the coefficients  $\aa_{r,c}$ we have found above 
into $D^{(1)}$ in \rf{dee} and try to 
formally perform the summation 
over  $r$ first, independently for each $Q_{a,a+c}$ 
term.
We get
\beqa
\label{D1 sum}
D^{(1)}
&=& 
2 \sum_{a=1}^L \sum_{c=1}^\infty 
f_c(\lambda)\  Q_{a,a+c} \ , \ \ \ \ \ \ \ \ \ \ 
f_c(\lambda)= \sum_{r=c}^\infty   
{\lambda^r \over (4\pi)^{2r}} 
 \ \aa_{r,c} \ . 
\eeqa 
Remarkably,  the series representation 
for the coefficients $f_c(\lambda)$ 
can then be summed up in terms of 
the standard hypergeometric functions (see   Appendix):
\beqa
\label{coeff-c}
f_c(\lambda) &=& 
\sum_{r=c}^\infty
 {\lambda^r \over (4\pi)^{2r}} \ 
{ (-1)^{r-c} \; 
\Gamma(2r-1) \Gamma(2r+1) 
\over 
\Gamma(r+1) \Gamma(r) \Gamma(r-c+1) \Gamma(r+c+1)}
\nonumber\\ &=&
 \left( {\lambda \over 4 \pi^2}\right)^c  
{\Gamma(c-\half) \over 4\sqrt \pi \ \Gamma(c+1)}
\;\ 
{}_2 F_{1} (c-\half, c+\half; 2 c + 1; - {\lambda \over \pi^2})\ . 
\eeqa
The coefficient in front of ${}_2 F_{1}$  is equal to 
$ {2\l^c\ov (4 \pi)^{2c}} \aa_{c,c}$ (cf. \rf{aBMN}).   
The $f_c$ go to 0 rapidly at large $c$, so we effectively have 
a spin chain with a short range interactions.

In general, the  hypergeometric functions 
$ {}_2 F_{1}(a_1,a_2; b_1; z)$ 
have a cut in the $z$ plane running from $1$ to $\infty$.
Note also that $y(z)= {}_2 F_{1} (c-\half, c+\half; 2 c + 1; z)$ 
solves the following differential  equation 
$z(1-z) y'' + (2c+1) ( 1-z) y' - (c^2 - {1\ov 4}) y =0$.

The resulting  coefficients 
$f_c(\l)$ are smooth positive  functions of $\l$
having   regular expansion at both  small $\l$ (see \rf{coeff-c}) 
and  
large $\l$ 
\beqa
\label{f:large}
f_c(\lambda)_{_{\l \to \infty} }
 &=& 
{\sqrt{\lambda} \over \pi^2 } 
\left[
{1 \over 4 c^2 - 1 } 
- 
{\pi^2 \over 4\lambda} 
\left(
 \ln {\lambda\ov \pi^2} +  1 - 2 H_{c-{1\over2}}
\right)
+ \OO({ 1 \ov \lambda^{2}}) \ , 
\right]
\eeqa
where the harmonic numbers $H_{p}$ are defined in Appendix.
The square root $f_c \to \sqrt \l$ asymptotics  of (\ref{f:large}) 
is related to  the cut structure of $ {}_2 F_{1}$.
The $\ln \l$  subleading terms are likely to be an artifact 
of our resummation procedure (they will be  absent in the 
explicit  $L=4$ example discussed below). 

One may be tempted to interpret the
 behavior of the coefficients $f_c$
as an indication of how  anomalous dimensions of particular operators 
should scale with $\l$ (one should remember of course that 
$D^{(1)}$ is only a part of the full dilatation operator in \rf{dee}). 
Their  $\sqrt \l$ asymptotics may  then
 seem to be in 
contradiction with the usual expectation that dimensions 
of generic operators corresponding to string modes should have  slower 
growth  with $\l$ -- they should scale as  
square root of the effective string tension, i.e. as 
$\sqrt[4]{\l}$  \ci{gkp}. 

A possible resolution 
of this  paradox 
is that the above resummation 
procedure leading to \rf{D1 sum},\rf{coeff-c} 
 is useful  only in the infinite chain 
$L \to \infty$ limit, i.e. it 
corresponds to the case when $D$ acts on  
 ``long'' operators. The latter  should be dual 
to semiclassical string modes  for which dimensions are 
 expected 
to grow  as string tension $\sim \sqrt \l$ \ci{gkp}.
Indeed, we have treated all $Q_{a,a+c}$ terms 
 as independent but 
for finite $L$ the terms with $c$ and $c+ m L$ are the same
because of the periodicity of the chain
(implied by cyclicity of the trace in the operators).
Also, under the sum over $a$ one has $Q_{a,a+c}= Q_{a,a+L-c}$, 
   i.e. 
\beqa
\label{Q:cycl}
\Pi_{c} = \Pi_{c+m L} = \Pi_{mL-c} \ , \ \ \ \ \ \ \ \ \ \ \ 
\Pi_c \equiv \sum_{a=1}^L Q_{a,a+c}\ , 
\eeqa
where $m$ is any positive integer number. 
Therefore, for finite $L$ 
the  sum over $c$ should, in fact, 
be restricted to run from $c = 1$ to $c = L-1$, 
\beqa
\label{D1:finite}
D^{(1)}
&=& 2 \sum_{a=1}^L \sum_{c=1}^\infty 
f_c(\lambda)\ Q_{a,a+c}
= 
\sum_{a=1}^L 
\sum_{c=1}^{L-1}
h_c(L,\lambda)\ Q_{a,a+c}\ ,  
\eeqa
or, equivalently, 
\beqa
\label{finite}
D^{(1)}= 2 \sum_{c=1}^{[L/2]} h_c(L,\lambda)\  \Pi_c \ .  
\eeqa 
The   new coefficients $h_c$
depend on both the  't
Hooft coupling $\lambda$ 
and the length of the chain $L$, 
\beqa
\label{c:finite}
h_c(L,\lambda) 
&\equiv& 
\sum_{m=0}^\infty 
\left[
f_{c+mL}(\lambda) 
+
f_{L-c+mL}(\lambda) 
\right]
\ 
,\hspace{2em}
\mbox{$
c = 1, ..., L-1;\  c \ne L/2
$}
\nonumber\\
h_{L/2}(L,\lambda) 
&\equiv& 
\sum_{m=0}^\infty 
f_{L/2 +mL}(\lambda)\ 
,\hspace{7.5em}
\mbox{if $L$ is even} \ . 
\eeqa
They  satisfy the  periodicity condition 
$h_c(L,\lambda)  = h_{L-c}(L,\lambda)$, 
reflecting the fact that for 
pairwise interactions it matters only 
 which sites participate  in the interaction. 
Explicitly, the coefficients appearing in 
(\ref{c:finite})
can be written using (\ref{coeff-c}) as
\beqa
\label{D1 fin}
h_c(L,\lambda) 
&=&
\sum_{m=0}^\infty 
\bigg[\ \left( {\lambda\ov   4 \pi^2} \right)^{c+m L} 
{
\Gamma(c+ mL -{1\over2})\over 
4 \sqrt{\pi}\  \Gamma(c + mL +1) }  
\nonumber\\&&\hspace{6em}
\times\  {}_2 F_1 (c + m L -\half, c + m L +\half; 2c + 2mL 
+1; - {\lambda \over \pi^2})\ \bigg] 
\nonumber\\&&\hspace{3em}
+\ \  (c \to L-c) 
\ . 
\eeqa
When $L \to \infty$ and $0 < c \ll L$, 
the only contribution to the sum (\ref{D1 fin}) 
comes from $m=0$ in the first term, and (\ref{D1 fin}) reduces 
to (\ref{coeff-c}). 
For finite $L$, 
  we  may 
 expand the hypergeometric functions in \rf{D1 fin}
 at large $\l$ as in (\ref{f:large})  and then 
do the sum over $m$. Ignoring the issue 
of convergence of the resulting strong-coupling expansion, 
that leads to the following  simple  result 
for the leading-order term 
\beqa
\label{D1 lrg} 
h_c (L,\lambda)_{_{\l \to \infty}}  =
{\sqrt{\lambda} \over 2\pi  L } 
\left[
{\sin {\pi \over L} \over \cos{\pi \over L} - \cos{2 \pi c\over L}}
+ \OO(\l^{-1})
\right]
.
\eeqa
In the $c \ll L $ limit  
(\ref{D1 lrg}) reduces   back  to the large $\l$ asymptotics \rf{f:large} 
of the $m=0$ term of the sum in \rf{D1 fin}.


It is possible  that  for finite $L$  the contributions 
of higher order $Q^n$ interaction terms 
 in  $D$ \rf{dee} may transform  this asymptotics into 
the expected $\sqrt[4]{\l}$  behavior. 
At the same time, one may  wonder if our basic assumption
about  the structure of $\D^{(1)}_{2r} $ in \rf{ru} actually applies
for  finite values of $L$ and all values of $r$.
After all, to fix the coefficients 
$\aa_{r,c}$ we used the condition of regularity of the 
scaling limit which assumes  that 
 $L \to \infty$.\foot{Possible  subtleties in
applying the 
 general expression for the dilatation operator to 
 operators with small length  $L$ were mentioned
 in \ci{bks} (footnote 18), \ci{beif} (footnote 4)  and 
  \ci{bei} (section 4.3).} 
 
Ignoring this cautionary note let us go ahead and 
apply the above relations to the 
first non-trivial small $L$ case --  $L=4$
(the operators with lengths 
 $L=2,3$  are BPS: the antisymmetric combinations vanish
because of trace cyclicity).
The non-BPS operator with $L=4$ 
is the level four descendant $K$ of the Konishi scalar operator
\beqa
\label{konishi}
K = \tr [X , Z]^2= 2 \, \tr (X Z X Z - X X Z Z) \ . 
\eeqa
The action of $D^{(1)}$ \rf{D1:finite}  on $K$ is determined by noting that
(see \rf{Q:cycl})  
\be
\label{Q-koni}
\Pi_1  K = \Pi_3 K = 6 K 
,\ \ \  \ \  \Pi_2  K = 0\ , \ \ \ \ \ \ \ 
D^{(1)} K = \gamma^{(1)}  K\ .
\ee
Using that the periodicity implies that 
$\Pi_1=\Pi_3=\Pi_5=\Pi_7=...$ one finds then directly from \rf{dee} 
\be \la{gam}
\gamma^{(1)} = \sum^\infty_{k=1} \sum_{p=1}^k
\bigg[ ({ \l \ov 16 \pi^2})^{2k}\   \aa_{2k, 2p-1}
+  ({ \l \ov 16 \pi^2})^{2k-1} \  \aa_{2k-1, 2p-1} \bigg] \ . \ee
Using \rf{fin} the  sums over $p$  can be found explicitly  
\be \la{siu}
 \sum_{p=1}^k  \aa_{2k, 2p-1} =-
  { 2^{4k-2} \G(4k-1) \ov \G(2k) \G(2k+1) } \ , \ \ \ \ \ \ \ 
  \sum_{p=1}^k  \aa_{2k-1, 2p-1} =  
{ 2^{4k-4} \G(4k-3) \ov \G(2k) \G(2k-1) } \ , \ee 
  and   finally we obtain the following surprisingly simple result 
  (cf. \rf{bin2}) 
\be  \la{konn}
\gamma^{(1)} = { 3\ov 2} \left( \sqrt{ 1 + {\lambda\ov \pi^2} } -1 \right)
\ . \ee 
Then for  small $\l$ we reproduce the previously known 
results  \ci{ans,bks} 
\be \la{koni}
\gamma^{(1)} =  3 {\lambda\over 4\pi^2} -  3 \left( {\lambda\over
4\pi^2}\right)^2 + 
{6} \left({ \lambda  \ov  4\pi^2}\right)^3  
-15  \left({ \lambda  \ov  4\pi^2}\right)^4 + 
 \OO(\lambda^5) \ ,  \ee
 while  in the  large $\lambda$ limit 
one  finds  
\beqa
\label{koni-lrg}
\gamma^{(1)} &=& 
{3 \over 2\pi } \sqrt{\lambda} - { 3\ov 2} +  { 3\pi \ov 2 \sqrt \l} 
+ \OO( { 1 \ov  \l^{3/2} } )
\ ,  \eeqa
where the  leading term agrees with 
 (\ref{D1 lrg}).
Note also  that in the $\l^3$ and $\l^4$  terms in \rf{koni} 
 we included only the contributions of $D^{(1)}_6$ and $D^{(1)}_8$,
 i.e.  the linear in $Q$ terms in \rf{ity} and
 \rf{nex}.
  The contributions of the $QQ$ terms in 
\rf{ity} and  \rf{nex}  
change   the 3-loop  and 4-loop coefficients in \rf{koni} from $6$
 to  ${21\ov 4}$ \ci{bks} and 
 from $-15$ to $-{705\ov 64}$ \ci{beif} (but see footnote 7).
Again, one    may hope   that a systematic account of
contributions of all   higher $\D^{(n)}$ terms  in \rf{stru} 
will change the strong-coupling asymptotics 
 of the dimension of the   Konishi operator from 
$\sqrt \l$ to  $ \sqrt[4]{\l} $.

\setcounter{equation}{0}
\section{Concluding  remarks}
\label{closing}

Inspired by recent work in \ci{bks,beif} and especially  \ci{serb}, 
in this paper we suggested to organize the dilatation operator as an 
expansion \rf{dee},\rf{stru} 
in powers of independent projection operators $Q_{a,b}$ 
\ci{serb,krt} 
at $L$  sites of spin chain 
\be 
\label{D-again}
D =D_0 +  \sum_{n=1}^\infty D^{(n)} 
\quad\mbox{ with }
D_0 = L 
, ~ \ \ \ 
D^{(1)} = \sum_{a=1}^L \sum_{c=1}^{L-1}  
h_c(L,\lambda)\ Q_{a,a+c} \ , \ ...
\ee
where $D^{(n)}$ are given by sums of products of $n$ $Q$'s 
at independent sites of the spin chain. 
We determined the coefficients in $D^{(1)}$ by 
demanding that its 
BMN-type scaling limit \ci{kru,krt}
be regular, 
and 
found that 
it 
admits a very simple 
representation \rf{D1 sum}, 
applicable at least in the large $L$ limit. 
This representation includes all orders in $\l$ 
and suggests that 
the corresponding anomalous dimensions should  grow  
as $\sqrt \l$ for large $\l$. 
  
A  natural extension of this work would be to try to  find 
the next term in the expansion (\ref{D-again}), namely 
\beqa
\label{D2-def}
D^{(2)}  =
\sum_{a=1}^L \sum_{c_1,c_2,c_3=1}^{L-1} 
\hspace{-1.5ex}{}'\hspace{1.5ex}
h_{c_1; c_2,c_3}(L,\lambda)\ 
Q_{a,a+c_1} Q_{a+c_2,a+c_2+c_3} \ 
. \eeqa
The prime on the sum here 
means that certain terms should be  omitted:
since 
 $Q_{a,b} Q_{b,c} + Q_{b,c} Q_{a,b} 
= Q_{a,b} + Q_{b,c} - Q_{a,c}$ the 
terms with 
$c_2=c_1$ and  $c_2=c_3-c_1$  
have already been included in $D^{(1)}$.
The contributions of higher $D^{(n)}$ terms should be 
crucial at  finite $L$, resolving, 
in particular,  the above-mentioned
contradiction between  the $\sqrt \l$ asymptotics 
of the coefficients in $D^{(1)}$ and the expected $\sqrt[4]{\l}$ scaling 
of dimensions of   operators corresponding 
to string modes.

The AdS/CFT duality suggests that $D$ should 
correspond to   an integrable spin   chain.
The simplest possibility 
could be that, by analogy with the Inozemtsev chain  \ci{serb}, 
the operator 
$D^{(1)}$ (with interaction coefficients given by \rf{D1 fin}) 
represents a Hamiltonian of an integrable 
spin 1/2 chain, while all  higher order
terms  $D^{(n)}$ are effectively determined
by $D^{(1)}$ through integrability.
This, however, is  unlikely in view of the low loop order results of 
\ci{bks,beif,bei} and the very recent paper 
\ci{new}  suggesting that the {\it full}  dilatation operator 
satisfying the requirements of integrability, 
BMN scaling and consistency  with 
gauge theory should be essentially unique.



\vskip -0.5cm

\section{Acknowledgments }
We are grateful to 
G. Arutyunov, 
N. Beisert, 
S. Frolov, 
M. Kruczenski,
A. Parnachev, A. Onishchenko, M. Staudacher  and K. Zarembo 
for useful 
discussions, 
e-mail correspondence on related issues
and comments on an earlier version of this paper. 
The work of A.R. was supported in part by 
NSF grants PHY99-73935 and PHY04-01667.
The work of A.T. was supported by 
DOE grant DE-FG02-91ER40690,
the INTAS contract 03-51-6346 and Wolfson award. 
We  also thank the Michigan Center
for Theoretical Physics for the  hospitality 
during the beginning of this work.

\vskip -0.5cm

\setcounter{section}{0}
\setcounter{equation}{0}
\setcounter{footnote}{0}
\appendix{Some useful relations and definitions}
\label{techni}

Here we summarize some useful formulae 
used in  the paper.
The usual binomial expansion is given by 
\beqa
\label{binom}
(1 + x)^n &=& 
\sum_{k=0}^\infty { \G(n +1)  \over \G(n-k +1)  \G(k+1) } \   x^k  \  
\eeqa
(for integer $n$ the series terminates at $k=n$).
For $n=\half$ (\ref{binom}) can be written as  
\beqa
\label{bin2}
(1 + x)^{1/2} &=& 
1 + 2 \sum_{k=1}^\infty {(-1)^{k-1} \Gamma(2k-1) \over \Gamma(k) \Gamma(k+1)} 
\left( {x \over 4} \right)^k
.
\eeqa
To transform the arguments of $\G$-functions one uses 
\beqa
\label{G rels}
\G(2z) = {2^{2z-1} \over \sqrt \pi} \G(z) \G(z+ \half)
\ ,\qquad
\G(z) \G(1-z) = {\pi \over \sin \pi z}\ . 
\eeqa
The hypergeometric functions  are given, 
within the radius of convergence, 
 by the series
\beqa
\label{hyper}
{}_p F_q (a_1, ... , a_p ; b_1, ... , b_q ; z)
= 
\sum_{k=0}^\infty
{
(a_1)_k ... (a_p)_k
\over
(b_1)_k ... (b_q)_k
}
{z^k\over k!} \ , \ \ \ \ \ \ \ \  (a)_k \equiv {\G(a+k) \ov \G(a) } \ . 
\eeqa
They reduce to simpler functions in particular cases; 
for example 
${}_2 F_1(-n, a; a, -z) = (1+z)^n$
as one can see by comparing (\ref{binom}) and (\ref{hyper}).
If $p=q+1$, ${}_p F_q (a_1, ... , a_p ; b_1, ... , b_q ; z)$
have a branch cut in the $z$-plane running from $z=1$ to $\infty$.
${}_p F_q $ 
satisfy second order differential equations, 
and by a change of variables 
one can relate the values of ${}_p F_q$  at $z$ and at $1/z$,
although for different arguments $a_i$ and $b_j$.
One can show that for large $z$ 
\beq
\label{hyper-lrg}
{}_2 F_1 (b-1, b ; 2b ; z)
=
{(-z)^{1-b} \Gamma(2b) \over \Gamma(b) \Gamma(b+1) }
\left[
1 + {b(b-1) \over (-z)} \left( 1 + \ln(-z) - 2 H_{b-{1\over2}} \right)
+ \OO(z^{-2})
\right]\ , 
\eeq
where the Harmonic numbers are given by 
\beqa
\label{Hn}
H_p \equiv H_p^{(1)}\ 
,\qquad
H_p^{(s)} = \zeta(s,1) - \zeta(s,p+1)\ , \ \ \ \ \ 
\zeta(s,p) \equiv  \sum_{k=0}^\infty (k+p)^{-s}\ , 
\eeqa
and for integer $p > 1$ one has 
$H_p^{(s)} = \sum_{k=1}^n k^{-s}$.
One finds that  $H_{b-{1\over2}} = \gamma_E + \psi(b)$, where  
$\psi(b)={\Gamma'(b)\ov \Gamma(b)}$,
and $\gamma_E = - \psi(1) \approx 0.577216$ is 
the Euler's constant.





\begin{thebibliography}{20}

\bi{gko}
F.~Gliozzi, J.~Scherk and D.~I.~Olive,
``Supersymmetry, Supergravity Theories And The Dual
Spinor Model,''
Nucl.\ Phys.\ B {\bf 122}, 253 (1977).
L.~Brink, J.~H.~Schwarz and J.~Scherk,
``Supersymmetric Yang-Mills Theories,''
Nucl.\ Phys.\ B {\bf 121}, 77 (1977).

\bi{fin}
 M.~F.~Sohnius and P.~C.~West,
``Conformal Invariance In N=4 Supersymmetric Yang-Mills
Theory,''
Phys.\ Lett.\ B {\bf 100}, 245 (1981).
 L.~Brink, O.~Lindgren and B.~E.~W.~Nilsson,
``The Ultraviolet Finiteness Of The N=4 Yang-Mills
Theory,''
Phys.\ Lett.\ B {\bf 123}, 323 (1983).
P.~S.~Howe, K.~S.~Stelle and P.~K.~Townsend,
``Miraculous Ultraviolet Cancellations In Supersymmetry
Made Manifest,''
Nucl.\ Phys.\ B {\bf 236}, 125 (1984).




\bibitem{mal}
J.~M.~Maldacena,
``The large $N$ limit of superconformal field
theories and supergravity,''
Adv.\ Theor.\ Math.\ Phys.\  {\bf 2}, 231
(1998)
[hep-th/9711200].
S.~S.~Gubser, I.~R.~Klebanov and
A.~M.~Polyakov,
``Gauge theory correlators from non-critical
string theory,''
Phys.\ Lett.\ B {\bf 428}, 105 (1998)
[hep-th/9802109].
E.~Witten,
``Anti-de Sitter space and holography,''
Adv.\ Theor.\ Math.\ Phys.\  {\bf 2}, 253
(1998)
[hep-th/9802150].


\bi {bmn}
D.~Berenstein, J.~M.~Maldacena and
H.~Nastase, ``Strings in flat space and
pp waves
from N =4 super Yang Mills,''
JHEP {\bf 0204}, 013 (2002)
[hep-th/0202021].



\bibitem{gkp}
S.~S.~Gubser, I.~R.~Klebanov and
A.~M.~Polyakov,
``A semi-classical limit of the
gauge/string correspondence,''
Nucl.\ Phys.\ B {\bf 636}, 99 (2002)
[hep-th/0204051].

  \bi{gro}
D.~J.~Gross, A.~Mikhailov and R.~Roiban,
``Operators with large R charge in N = 4 Yang-Mills
theory,''
Annals Phys.\  {\bf 301}, 31 (2002)
[hep-th/0205066].
``A calculation of the plane wave string Hamiltonian
from N = 4
super-Yang-Mills theory,''
JHEP {\bf 0305}, 025 (2003)
[hep-th/0208231].

\bi {sz}
A.~Santambrogio and D.~Zanon,
``Exact anomalous dimensions of N = 4
Yang-Mills operators with
large R
charge,''
Phys.\ Lett.\ B {\bf 545}, 425 (2002)
[hep-th/0206079].



\bibitem{ft1}
S.~Frolov and A.~A.~Tseytlin,
``Semiclassical quantization of
rotating superstring in \adss,''
JHEP {\bf 0206}, 007 (2002)
[hep-th/0204226].

 \bi{rus}
 J.~G.~Russo,
``Anomalous dimensions in gauge theories from rotating
strings in  AdS(5) x S(5),''
JHEP {\bf 0206}, 038 (2002)
[hep-th/0205244].

 \bi{mina}
 J.~A.~Minahan,
``Circular semiclassical string solutions on $AdS_5 \times S^5$,''
Nucl.\ Phys.\ B {\bf 648}, 203 (2003)
[hep-th/0209047].

 \bibitem{mz1}
J.~A.~Minahan and K.~Zarembo,
``The Bethe-ansatz for N = 4 super
Yang-Mills,''
JHEP {\bf 0303}, 013 (2003)
[hep-th/0212208].


\bibitem{ft2}
S.~Frolov and A.~A.~Tseytlin,
``Multi-spin string solutions in
\adss,''
Nucl.\ Phys.\ B {\bf 668}, 77 (2003)
[hep-th/0304255].

\bibitem{bmsz}
N.~Beisert, J.~A.~Minahan, M.~Staudacher
and K.~Zarembo,
``Stringing spins and spinning strings,''
JHEP {\bf 0309}, 010 (2003)
[hep-th/0306139].

\bibitem{ft4}
S.~Frolov and A.~A.~Tseytlin,
``Rotating string solutions: AdS/CFT duality in
non-supersymmetric
sectors,''
Phys.\ Lett.\ B {\bf 570}, 96 (2003)
[hep-th/0306143].


\bibitem{bfst}
N.~Beisert, S.~Frolov, M.~Staudacher and
A.~A.~Tseytlin,
``Precision spectroscopy of AdS/CFT,''
JHEP {\bf 0310}, 037 (2003)
[hep-th/0308117].


\bi{as}
G.~Arutyunov and M.~Staudacher,
``Matching higher conserved charges for strings and spins,''
JHEP {\bf 0403}, 004 (2004)
[hep-th/0310182].
``Two-loop commuting charges and the string / gauge duality,''
hep-th/0403077.



\bibitem{emz}
J.~Engquist, J.~A.~Minahan and
K.~Zarembo,
``Yang-Mills duals for semiclassical
strings
on \adss,''
JHEP {\bf 0311}, 063 (2003)
[hep-th/0310188].


\bi{kru}
M.~Kruczenski,
``Spin chains and string theory,''
hep-th/0311203.

\bi{serb}
D.~Serban and M.~Staudacher,
``Planar N = 4 gauge theory and the Inozemtsev
long range spin chain,''
hep-th/0401057.

\bibitem{kmmz}
V.~A.~Kazakov, A.~Marshakov, J.~A.~Minahan and
K.~Zarembo,
``Classical/quantum integrability in
AdS/CFT,''
hep-th/0402207.

\bibitem{krt}
M.~Kruczenski, A.~V.~Ryzhov and A.~A.~Tseytlin,
 ``Large spin limit of \adss  string theory and low energy
 expansion of
ferromagnetic spin chains,''
hep-th/0403120.

\bi{kot}
A.~V.~Kotikov, L.~N.~Lipatov and V.~N.~Velizhanin,
``Anomalous dimensions of Wilson operators in N = 4 SYM theory,''
Phys.\ Lett.\ B {\bf 557}, 114 (2003)
[hep-ph/0301021].
A.~V.~Kotikov, L.~N.~Lipatov, A.~I.~Onishchenko and V.~N.~Velizhanin,
``Three-loop universal anomalous dimension of the Wilson operators in N
= 4
SUSY Yang-Mills model,''
hep-th/0404092.



\bibitem{bks}
N.~Beisert, C.~Kristjansen and M.~Staudacher,
``The dilatation operator of N = 4 super Yang-Mills
theory,''
Nucl.\ Phys.\ B {\bf 664}, 131 (2003)
[hep-th/0303060].

\bibitem{beif}
N.~Beisert,
``Higher loops, integrability and the
near BMN
limit,''
JHEP {\bf 0309}, 062 (2003)
[hep-th/0308074].

\bibitem{bei}
N.~Beisert,
``The su(2$|$3) dynamic spin chain,''
hep-th/0310252.

 \bi{plefka}
 T.~Klose and J.~Plefka,
``On the integrability of large N plane-wave matrix
theory,''
Nucl.\ Phys.\ B {\bf 679}, 127 (2004)
[hep-th/0310232].

 \bi{sha}
 F.D.M. Haldane, ``Exact Jastrow-Gutzwiller Resonating
 Valence Bond Ground State Of The Spin 1/2
 Antiferromagnetic Heisenberg Chain With 1/$R^2$
  Exchange'', Phys.\ Rev.\ Lett. {\bf 60}, 635  (1988);
   S. Shastry,  ``Exact Solution Of An S = 1/2
   Heisenberg Antiferromagnetic Chain With Long Ranged
   Interactions'', Phys.\ Rev.\ Lett. {\bf 60}, 639
   (1988).

  \bi{ino}
   V.~I.~Inozemtsev,
  ``On The Connection Between The One-Dimensional S =
  1/2 Heisenberg Chain And
Haldane Shastry Model,''
J. Stat. Phys. {\bf 50}, 1143 (1990);
``Integrable Heisenberg-van Vleck chains with variable
range exchange,''
Phys.\ Part.\ Nucl.\  {\bf 34}, 166 (2003)
[Fiz.\ Elem.\ Chast.\ Atom.\ Yadra {\bf 34}, 332 (2003)]
[hep-th/0201001].


\bi{fra}
E.~H.~Fradkin,
``Field Theories Of Condensed Matter Systems,''
 Redwood City, USA: Addison-Wesley (1991) 350 p. (Frontiers in
physics, 82).


\bi{bebmn}
N.~Beisert,
``BMN operators and superconformal symmetry,''
Nucl.\ Phys.\ B {\bf 659}, 79 (2003)
[hep-th/0211032].

\bi{ans}
D.~Anselmi,
``The N = 4 quantum conformal algebra,''
Nucl.\ Phys.\ B {\bf 541}, 369 (1999)
[hep-th/9809192].

\bi{new}
N. Beisert, V. Dippel and  M. Staudacher, 
``A Novel Long Range Spin Chain and Planar N=4 Super Yang-Mills'', 
hep-th/0405001.



\end{thebibliography}
\end{document}